# On the Validity of the Assumption of Local Equilibrium in Non-Equilibrium Thermodynamics

**Arieh Ben-Naim**

**Department of Physical Chemistry**

**The Hebrew University of Jerusalem**

**Givat Ram, Jerusalem 91904**

**Israel**

**Abstract**

The extension of equilibrium thermodynamics to non-equilibrium systems is based on the assumption of *"local equilibrium,"* followed by the assumption that an entropy-density function may be defined, and that this entropy-density would have the same functional dependence on the *local densities*, $(u, v, n)$, as the total entropy $S$ on the variables $(U, V, N)$.

In this article we question the validity of all the assumptions made in the theory of non-equilibrium thermodynamics. We find that the assumption of "local equilibrium" is ill-founded, that it is not clear how to define an entropy-density function, and it is far from clear that such a function, if it exists at all could be integrated to obtain the so-called "entropy production" of the entire system.

These findings shed serious doubt on the validity of applying thermodynamic theory to systems far from equilibrium systems.

## 1. Introduction

Non-equilibrium thermodynamics is founded on the assumption of *local equilibrium*.[1-6] Typically, this assumption states that:[1]

*"It will now be assumed that, although the total system is not in equilibrium, there exists within small elements a state of 'local' equilibrium for which the local entropy s is the same function of $u, v$ and $c_k$ as in real equilibrium."*

No justification is provided for this assumption, nor a proof that such a "local entropy" would have the "same function of $u, v$ and $c_k$ as in real equilibrium."

Most textbooks on non-equilibrium thermodynamics starts with the reasonable assumption that in such system the intensive variables such as temperature $T$, pressure $P$, and chemical potential $\mu$ may be defined operationally, in each small element of volume $dV$ of the system.

Thus, one writes

$$T(\mathbf{R},t), \ P(\mathbf{R},t), \ \mu(\mathbf{R},t) \qquad (1.1)$$

where $\mathbf{R}$ is the locational vector of a point in the system, and $t$ is the time. One can further assume that the density $\rho(\mathbf{R},t)$ is defined locally at point $\mathbf{R}$ and integrate over the entire volume to obtain the total number of particles $N$

$$N = \int \rho(\mathbf{R},t) d\mathbf{R} \qquad (1.2)$$

Similarly, one can define densities $\rho_k(\mathbf{R},t)$ for each component, k, of the system.

One can also defines the internal energy per unit of volume $u(\mathbf{R},t)$. It is not clear however, how to integrate $u(\mathbf{R},t)$ over the entire volume of the system to obtain the total internal energy of the system. While this may be done exactly for ideal gases, i.e. when the total energy of the system is the sum of all the kinetic (as well as internal) energies of all the particles in the system, it is not clear how to do the same for systems having interacting particles for which one defines the integral

$$U = \int u(\mathbf{R},t) d\mathbf{R} \qquad (1.3)$$

Here, the integration is essentially the sum over all the $u(\mathbf{R}, t)$, each of which is defined in a small cell $d\mathbf{R}(dxdydz)$, or $dV$ in the system, neglecting the interaction energies between the different cells. When there are interactions between the particle, it is not clear how to account for these interactions in the very definition of the local energy, $u(\mathbf{R}, t)$. Similarly, it is far from clear whether one can write the entropy of the entire system as an integral of the form:

$$S = \int s(\mathbf{R}, t) d\mathbf{R} \qquad (1.4)$$

The most important and unjustified assumption is related to the definition of the local entropy $s(\mathbf{R}, t)$. One assumes that the local entropy function, $s(u, v, n)$ is the same as the function $S(U, V, N)$, i.e. $s$ is the same function of the local energy, volume, and number of particles of each element of volume.

This assumption may be justified for ideal gas when the distribution of locations and velocities is meaningful for each element of volume in the system. To the best of the author's knowledge this assumption has never been justified for systems of interacting particles. The main difficulty is that for such systems there are correlations between the elements of volumes, which, in turn, adds mutual information to the Shannon measure of information (SMI) of the system,[7-11] hence also on the entropy of the system. When these correlations are large, it is not clear whether one can define the local entropy, and therefore the validity of the integral is questionable.

Once one makes such an assumption, one writes the changes in the entropy of the entire system as:

$$dS = d_e S + d_i S \qquad (1.5)$$

where $d_e S$ is the entropy change due to the heat exchange between the system and its surrounding, and $d_i S$ is the entropy produced in the system. For an isolated system $d_e S = 0$, and all the entropy change is due to $d_i S$. The latter is further written as

$$\frac{d_i S}{dt} = \int \sigma d\mathbf{R} \qquad (1.6)$$

where $\sigma(\mathbf{R}, t)$ is referred to the *local* entropy production

$$\sigma(\mathbf{R}, t) = \frac{d_i s}{dt} \geq 0 \qquad (1.7)$$

Thus, for an isolated system one has a local entropy production which is a function of time, and after integration one obtains also the total change of the entropy of the system as a function of time. Since the quantity $\sigma$ is defined in terms of the local entropy function $s(\mathbf{R}, t)$, and since $s(\mathbf{R}, t)$ is not a well-defined quantity, one should doubt the whole theory based on the assumption of local equilibrium.

In the following sections, we will show that:

1. It is *not clear* how to *define* a local entropy density function $s(\mathbf{R}, t)$.
2. It is *almost certain* that if one could have defined an entropy density, it would not have the same functional dependence on the local quantities $(u, v, n)$ as the entropy of the whole system depends on the macroscopic variables $(U, V, N)$.
3. It is *not true* that such an entropy-density function $s(R, t)$ could be integrated to obtain the entropy of the entire system.

## 2. How to define a local entropy-density function?

When a system is at equilibrium, the entropy is a homogenous function of its extensive variables, i.e. for any $\lambda > 0$ one can write

$$S(\lambda U, \lambda V, \lambda N_1, \cdots, \lambda N_c) = \lambda S(U, V, N_1, \cdots, N_c) \qquad (2.1)$$

One can use this property to *define* a local entropy-density simply by choosing $\lambda = V^{-1}$, i.e.

$$s = S/V = S\left(\frac{U}{V}, \frac{N_1}{V}, \cdots, \frac{N_c}{V}\right) \qquad (2.2)$$

where $N_i/V$ is the number density of the $i$th component. Clearly, at equilibrium, $s$ has the same value in any element of volume $dV$ at a point $\mathbf{R}$ in the same system. Therefore, one can integrate over the volume of the system to obtain the total entropy of the system

$$\int s(\mathbf{R}) dV = s(\mathbf{R}) V = S \qquad (2.3)$$

This procedure would not work for a system which is not at equilibrium. The reason is that in a non-equilibrium system we do not have the entropy of the entire system to begin with. Therefore, we cannot define the entropy-density as in Eq. (2.2).

For a non-equilibrium system we have to define first the local entropy-density then integrate to obtain the total entropy of the system.

Let us examine whether this can be done by using the following three definitions of entropy.

(i) *Using Clausius' definition*

Clausius originally defined a small change of entropy $dS$ due to a small addition of heat $dQ$ to a system at constant temperature. Implicit in this definition is the assumption that while we transfer $dQ$ the system's temperature does not change. This assumption is valid for a *small* addition of heat $dQ$, to a *macroscopic* system at a constant temperature. It would not work for a very small system. Even if one can define the local temperature $T(\mathbf{R})$, the addition of $dQ$ to an infinitesimal volume $dV$ would cause a very large change in the local temperature.

(ii) *Using Boltzmann's definition*

It is well-known that starting with the Boltzmann definition of entropy $S = k_B \ln W$ ($k_B$ being the Boltzmann constant, and $W$ the number of accessible microscopic states of the system), one can derive an explicitly entropy function $S(U,V,N)$ for an ideal gas (of one component)

$$S(U,V,N) = N k_B \ln\left[\frac{V}{N}\left(\frac{4\pi m U}{3h^2 N}\right)^{3/2}\right] + \frac{5}{2} N k_B \qquad (2.4)$$

This is the Sackur-Tetrode equation[14,15] (where $m$ is the mass of the particles, $h$ is the Planck constant).

Clearly, the function $S(U,V,N)$ has the property of being a homogenous function of first order. However, in the derivation of the Sackur-Tetrode equation one assumes that the number of particles $N$ is large enough so that the Stirling approximation may be applied to $N!$, i.e.

$$\ln N! \approx N \ln N - N \qquad (2.5)$$

For small elements of volumes the number of particles in each element might be so small that this approximation no longer applies.

This is certainly true for infinitesimal volumes in which the number of particles will either be zero or one. If one follows the derivation of Sackur-Tetrode equation and stops before the introduction of the Stirling approximation, one gets the equation

$$S \sim k_B \left[ N \ln \left[ V \left( \frac{2\pi e m k_B T}{n^2} \right)^{3/2} \right] - k_B \ln N! \right] \quad (2.6)$$

Clearly, for $N$ of the order of 1 this "entropy" will not be a homogenous function of order one is N. Therefore, an entropy-density function will not have the same dependence on $(u, v, n)$ as the entropy function $S(U, V, N)$.

*(iii)    Using the SMI-based definition*

The definition of entropy based on the Shannon measure of Information (SMI)[12] leads to the Sackur-Tetrode equation for an ideal gas of simple particles[7-11]. Therefore, the conclusion reached based on the Boltzmann definition will be the same as that reached by the definition based on the SMI.

Thus, we can conclude that whatever definition one used for the entropy of a macroscopic system, it will not apply to an infinitesimally small system. Therefore, one cannot assume that the local entropy density will have the same dependence on $(u, v, n)$ as the function $S(U, V, N)$.

### 3. The problem of integrability of the local entropy density

Suppose that somehow one could define a local entropy-density $s(\mathbf{R}, t)$, such that $s(\mathbf{R}, t)dV$ is the entropy of a small element of volume $dV$, at time T. The question is whether one may integrate such a function to obtain the entropy of the entire system, $S(t)$, i.e.

$$S(t) = \int_V s(\mathbf{R}, t) dV \quad (3.1)$$

We examine the question of integrability by starting with two independent systems A and B, each being a macroscopic system at equilibrium. In such a case, the additivity property of the entropy applies, i.e.

$$S(A + B) = S(A) + S(B) \quad (3.2)$$

Such an additivity applies to any number of independent systems, i.e.

$$S(total) = \sum_{i=1}^{c} S(i) \qquad (3.3)$$

where $S(i)$ is the equilibrium thermodynamic entropy of system $I$, Figure 1.

Next, we bring the systems into interacting-contact. We imagine that each system is an isolated system with a well-defined entropy $S(i)$. We assume that the walls are impermeable to heat, to volume, and to particles but they are so thin that particles in $i$ can interact with particles in $j$, where $i$ and $j$ are adjacent systems, Figure 2.

Such idealized walls do not exist, and we shall soon remove these. However, assuming the existence of such walls, we could write the total entropy of all the $c$-systems as

$$S(Total) = \sum_{i=1}^{c} S(i) + \sum_{\substack{i=1 \\ i<j}}^{c} \sum_{j=1}^{c} S_{in}(i,j) \qquad (3.4)$$

where, as before $S(i)$ is the (macroscopic) entropy of the $i$th system, and $S_{in}(i,j)$ is the entropy due to the intramolecular interaction between particles in $j$, and particles in $i$. These entropies of interactions can be cast in the form of mutual information between any pairs of systems.[7-11]

When all the systems are macroscopic one can neglect all the interaction entropies between the systems. This assumption follows from the fact that only particles on the surface of $i$ can interact with particles on the surface of $j$. Since the number of particles on the surface of each system is negligible (for macroscopic systems) compared with the number of interacting particles within each system we can neglect the mutual information between any pair of system (hence, the entropy of interaction) compared to the mutual information due to intermolecular interaction within each system. Such an approximation will preserve the additivity property (3.3) in spite of the existence of interactions between the systems.

Clearly, this approximation becomes less and less valid when the systems become smaller. In the limit when the "systems" are infinitesimally small, the interactions between particles in different elements of volume ("system") will dominate the entropy due to interactions in the entire system.

For non-ideal gases, when we can neglect high order correlation functions, the mutual information (MI) in the entire system (hence, the entropy due to intermolecular interactions) has the form

$$MI = -\frac{N(N-1)}{2}\int P(\boldsymbol{R}_1, \boldsymbol{R}_2)\log[g(\boldsymbol{R}_1, \boldsymbol{R}_2)]d\boldsymbol{R}_1, \boldsymbol{R}_2 \qquad (3.5)$$

where $P(\boldsymbol{R}_1, \boldsymbol{R}_2)d\boldsymbol{R}_1 d\boldsymbol{R}_2$ is the probability of finding a specific particle in $d\boldsymbol{R}_1$ at $\boldsymbol{R}_1$, and another specific particle in $d\boldsymbol{R}_2$ at $\boldsymbol{R}_2$.

It is not clear how to divide this total mutual information into a sum (or an integral) over all the elements of volume $dV$. Specifically, when each element of volume can accommodate almost one particle, hence, such an element of volume would not have any mutual information (hence, no entropy due to interactions).

The situation is far more serious in a condensed phase when higher order correlations become significant. As is well-known even when intermolecular interactions are strictly pairwise, higher order correlations are not negligible. This in turn means that instead of (3.5), we must include mutual information between all particles in the system, i.e.

$$MI = -\int P(\boldsymbol{R}^N)\log[g(\boldsymbol{R}^N)]dR^N \qquad (3.6)$$

Where

$$g(\boldsymbol{R}^N) = \frac{P(R^N)}{\prod_{i=1}^{N} P(\boldsymbol{R}_i)} \qquad (3.7)$$

This mutual information is a property of the entire system, and cannot be "divided" into contributions due to small, or infinitesimal elements of volume in the system.

## 4. Conclusion

To the best of our knowledge the assumption of "local equilibrium," followed by the definition of entropy-density function in a system far from equilibrium has never been founded either within thermodynamics, or from statistical mechanical arguments.

In this article we argued that the very assumption of "local-equilibrium" cannot be justified for systems far from equilibrium.

Furthermore, we argued that an entropy density function, even if it could be defined would not have the same functional dependence on the local energy, volume, and number of particles $(u, v, n)$ as the entropy of a macroscopic system at equilibrium $S(U, V, N)$. We also showed that for condensed systems, where intramolecular interactions are significant, one cannot derive an entropy function for the whole system by integrating over the *local* entropy-density function.

These findings shed serious doubts on the thermodynamics of systems far from equilibrium.

Appendix A.

In this Appendix a few quotations from the literature will be presented:

1. Glansdorff and Prigogine (1971) on page 12 start with the division of $dS$ into two terms

$$dS = d_i S + d_e S$$

where $d_e S$ is the change in entropy due to the flow of heat (i.e. $dS = \frac{dQ}{T}$), and $d_i S$ is the "entropy production due to changes inside the system."

Then, they write the equation

$$\frac{d_i S}{dt} = \int \sigma[S] dV \geq 0$$

Calling $\sigma[S]$ the "entropy source," the entropy production per unit time, and volume"

Then on page 14:

*"This will be the case when there exists within each small mass element of the medium a state of local equilibrium for which the local entropy s is the same function of the local macroscopic variables as at equilibrium state. This assumption of local equilibrium is not in contradiction with the fact that the system as a whole is out of equilibrium."*

No justification is provided for the assumption of local equilibrium.

2. de Groot and Mazur (1962) start from the formulation of the Second Law as

$$dS \geq 0 \text{ (for an adiabatically insulated system)}$$

Then introduce the entropy $s$ per unit mass defined by the equation

$$S = \int_V \rho s dV$$

where $\rho$ is the mass density (mass per unit volume) $\rho_k$ for the component $k$. they they write:

*"It will now be assumed that, although the total system is not in equilibrium, there exists within small elements a state of "local" equilibrium, for which the local entropy s is the same function (14) of $u, v,$ and $c_k$ as in real equilibrium."*

Thus, the assumption is made that the "local entropy" $s$ is the same function of $(u, v, c_k)$ as in real equilibrium system.

No justification for this "hypothesis of *local* equilibrium is provided. They claim that this can be justified by virtue of the validity of the conclusions derived from it.

3. Callen (1985), in Chapter 14 on "Irreversible Thermodynamics"

*"One problem that immediately arises is that of defining entropy in a nonequilibrium system. This problem is solved in a formal manner as follows:*

*To any infinitesimal region we associate a local entropy $S(X_0, X_1, ...)$, where, by definition, the functional dependence of S on the local extensive parameters $X_0, X_1, ...$ is taken to be identical to the dependence in equilibrium.*

*Again, the local intensive parameter $F_k$ is taken to be the same function of the local extensive parameters as it would be in equilibrium. It is because of this convention, incidentally, that we can speak of the temperature varying continuously in a bar, despite the fact that thermostatics implies the existence of temperature only in equilibrium systems.*

*The rate of local production of entropy is equal to the entropy leaving the region, plus the rate of increase of entropy within the region."*

I doubt that the introduction of the "local equilibrium" assumption "solves" the problem.

4. Kondepudi and Prigogine (1999), in Modern Thermodynamics, on pages 6-7 introduce the energy density $u[T(x), n(x)]$ defined as the "*internal energy per unit volume*. Then the total internal energy U is obtained as

$$U = \int_V u[T(x), n_k(x)]dV$$

where the integration is carried over the entire volume of the system.

$n_k(x)$ is the number of moles of the *k*th component per unit volume at point *x*.

Then, they write

$$N = \int_V n_R(x)dV$$

This should be $N_k$, the total number of molecules of type *k*. While $N_k$ can be defined as a sum or integral over all the elements of volumes in the system, the equation for the energy is valid only when the local $T(x)$ is well-defined at each point *x*, and that neglects interaction between the different elements of volume. This assumption is not discussed and not justified. Similarly, they define the entropy density $s[T(x), n(x)]$, and relate it to the total entropy of the system

$$S = \int_V s[T(x), n(x)]dV$$

If there are interactions between the particles, it is not clear how one *defines* the entropy density, and it is far from clear how one *defines* the entropy density, and it is far from clear that the entropy of the system can be written as an integral over the entropy densities.

It is important to note that the authors *do not* define $s[T(x), n(x)]$! They only say that "Similarly, an entropy density, $s(T, n_k)$ can be defined" without actually defining it. Clearly, the integral above does not *define* either the entropy of the system, or the entropy density $s(T, n_k)$.

On page 7, the authors comment:

*"In texts on classical thermodynamics, when it is sometimes stated that the entropy of a non-equilibrium system is not defined, it is meant that S is not a function of the variables U, V, and N. If the temperature of the system is locally well defined, then indeed the entropy of a non-equilibrium system can be defined in terms of an entropy density, as in (1.2.3)."*

This comment is doubly misleading. First, when people say that entropy of a non- equilibrium system is not defined, they mean *it is not defined*. Second, even if the temperature of the system is locally well defined (I doubt that this can be defined at any arbitrary small element of volume), it does not follow that equation for the entropy above, defines the entropy of the system in terms of an undefined entropy density.

5. In a more recent textbook by Kondepudi (2008), we find the explicit assumption made about the *local equilibrium*.

"*The basis of the modern approach is the notion of local equilibrium. For a very large class of systems that are not in thermodynamic equilibrium, thermodynamic quantities such as temperature, concentration, pressure, internal energy remain well-defined concepts locally, i.e. one could meaningfully formulate a thermodynamic description of a system in which intensive variables such as temperature and pressure are well defined in each elemental volume, and extensive variables such as entropy and internal energy are replaced by their corresponding densities. Thermodynamic variables can thus be functions of position and time. This is the assumption of local equilibrium.*"

Although I agree that local intensive quantities such as temperature, pressure, and densities may be defined operationally, I doubt that this can be done for either the internal energy or the entropy. No justification for this assumption is provided. Although the author admits that his assumption might not be a good approximation for some system, he does not explain for which systems the approximation is valid.

6. Kreuzer (1981), on page 3 of his book writes:

"*To explicitly state the conditions under which the assumption of local equilibrium is valid, the methods of nonequilibrium statistical mechanics are required. So far, this has been done rigorously and explicitly only for a dilute gas.*"

Indeed, if the system is an ideal gas, or a real gas but with negligible interaction energy, then one can express the total energy of the system as a sum or an integral over all infinitesimal values.

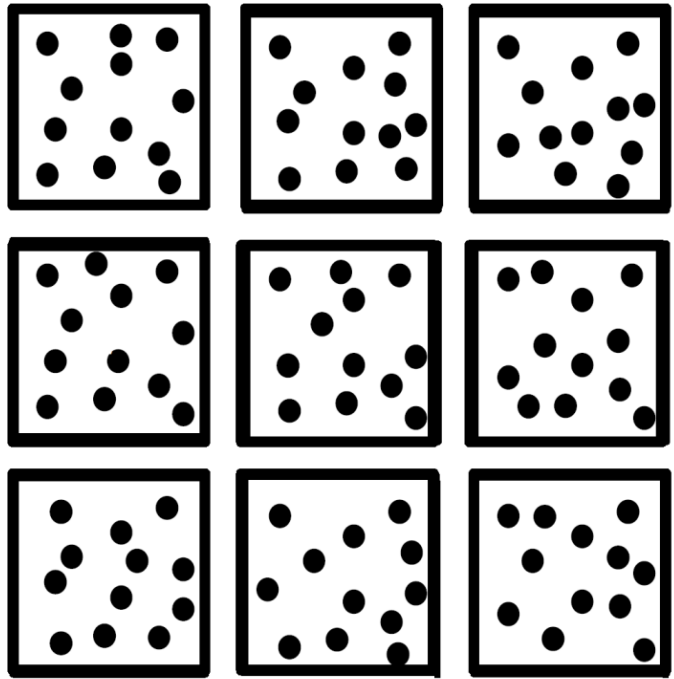

Figure 1. Six isolated systems,
Each having the same *energy, volume and number of particles*, at equilibrium

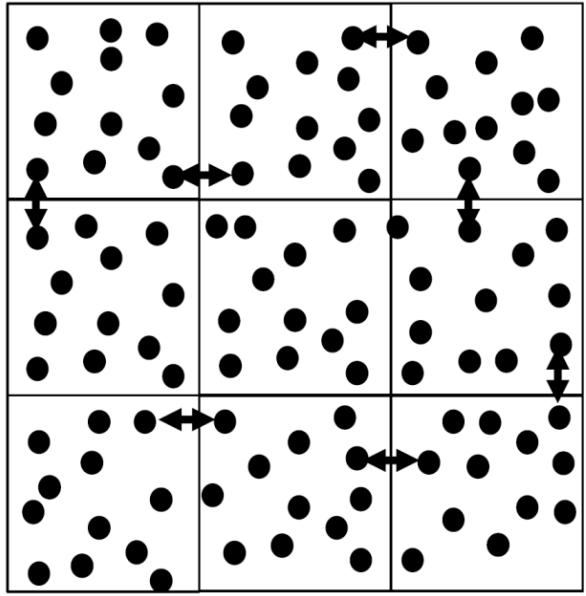

**Figure 2.** Six isolated systems,
each having the same *energy, volume and number of particles*,
but particles can interact across the walls.
Some of the interactions are shown by a double arrow